\documentclass[12pt]{article}
\usepackage{epsfig}
\usepackage{graphicx}
\usepackage{psfig}
\def\lsim{\mathrel{\rlap{\lower4pt\hbox{\hskip1pt$\sim$}}
    \raise1pt\hbox{$<$}}}         
\def\gsim{\mathrel{\rlap{\lower4pt\hbox{\hskip1pt$\sim$}}
    \raise1pt\hbox{$>$}}}         
\title{Neutrino Physics: an Update}
\author{Wick C. Haxton$^a$ and Barry R. Holstein$^{a,b}$\\
$^a$ Institute for Nuclear Theory\\
and Department of Physics\\
University of Washington\\
Seattle, WA  98195\\
$^b$ Department of Physics-LGRT\\
University of Massachusetts\\
Amherst, MA  01003}
\begin{document}
\begin{titlepage}
\maketitle
\begin{abstract}
We update our recent didactic survey of neutrino physics,
including new results from the Sudbury Neutrino Observatory
and KamLAND experiments, and recent constraints from WMAP
and other cosmological probes.
\end{abstract}
\end{titlepage}
\section{Introduction}
Several years ago, we authored a paper in this journal entitled
``Neutrino Physics,'' hereafter called I\cite{nui}, in order to
encourage inclusion of material involving neutrinos into the
introductory curriculum. We noted at the time that neutrinos, with
new experiments about to provide first data, might continue to be
a popular press item. This prediction has proven true:
\begin{itemize}
\item [i)] Ray Davis, the founder of the field of experimental solar neutrino
physics, shared the 2002 Nobel Prize in physics with
Mastoshi Koshiba, who led the Kamioka solar neutrino
experiment, and Riccardo Giacconi \cite{dav}.
\item [ii)] Results from the Sudbury Neutrino
Observatory (SNO) resolved the solar neutrino puzzle, showing
that approximately two-thirds of these neutrinos oscillate into
other flavors before reaching earth \cite{sno1,sno2}.
\item [iii)] The KamLAND experiment, in which antineutrinos from
Japanese power reactors were detected, confirmed the SNO results
and further narrow the allowed range of neutrino mass differences
\cite{kam}.
\item [iv)] The Wilkinson Microwave Anisotropy Probe (WMAP) measured
subtle temperature differences within the oldest light in the
universe, from the epoch when atoms first formed 380,000 years
after the Big Bang \cite{wmap}.  When combined with the results of
large scale structure studies \cite{2d}, a new bound on the sums
of the neutrino masses is obtained.
\end{itemize}
In addition there has been published (and disputed) evidence for the
existence of neutrinoless double beta decay which, if confirmed,
would show that one of the standard model's most important
symmetries, the conservation of lepton number, is violated \cite{dbd}.
Thus we decided to bring our early paper up to date by explaining
here the importance and implications of the new results.

In Section 2 we present a much abbreviated summary of the material
presented in I.  In Section 3 we discuss the SNO results and how
they resolved the puzzling discrepancies Ray Davis first uncovered
30 years ago. In Section 4 we discuss KamLAND, the first
terrestrial experimental to achieve the sensitivity to the small
neutrino mass differences relevant to solar neutrino experiments.
In Section 5 we describe marvelous new cosmological probes of
large scale structure and of the time when atoms were first
formed, and why the new data may soon challenge recent double beta
decay claims. We summarize where we stand in neutrino physics --
including the new discoveries that may be soon be within reach --
in Section 6.

\section{The Two-Minute Review}

In I we summarized the basics of neutrino physics, including
neutrino history, properties, and implications for contemporary
physics.  For the purposes of the present work, we note that in
the so-called standard model of particle/nuclear
physics\cite{dgh}, which is consistent with nearly all present
experimental information, there exist three massless neutrino
types---$\nu_e,\nu_\mu,\nu_\tau$---which are produced with purely
left-handed helicity in weak interaction processes.  If the
neutrino were shown to have a nonvanishing mass, it would be the
first clear failure of this 30-year-old standard model and the
first proof of the existence of the ``particle dark matter'' that
appears necessary to explain the structure and expansion of our
universe.  Nonzero neutrino masses were suggested by the results
of experiments measuring the flux of solar neutrinos, which are
produced as a byproduct of the thermonuclear reactions occurring
in the high-temperature core of our sun \cite{sol}.  Additional
strong evidence comes from the study of atmospheric neutrinos,
which are produced when high energy cosmic rays collide with the
upper atmosphere, producing pions and other particles that then
decay into neutrinos \cite{atm}.  The largest of the various
atmospheric neutrino experiments is SuperKamiokande, a detector in
a mine in the Japanese alps that contains 50,000 tons of
ultra-pure water.  SuperKamiokande's precise data show that the
flux of muon-type atmospheric neutrinos arriving from the opposite
side of the earth, which have travelled a long distance to reach
the detector, is depleted.

Both the solar and atmospheric neutrino results can be explained
quantitatively if neutrinos are massive and if the mass
eigenstates are not coincident with the weak interaction
eigenstates $\nu_e,\nu_\mu,\nu_\tau$, i.e., if the neutrinos
produced in weak interactions are combinations of the various mass
eigenstates.  This is exactly what is known to occur in the
analogous case of the quarks\cite{dgh}.  The relation between the
neutrino mass and weak-interaction eigenstates is described by a
unitary matrix: for three neutrinos, the mass and weak-interaction
eigenstates can be viewed as two distinct 3D coordinate systems.
The unitary matrix specifies the three rotations describing the
orientation of one coordinate system relative to the other.

The reduced flux of atmospheric muon neutrinos (see Figure 1) is
quantitatively explained by $\nu_\mu \rightarrow \nu_\tau$
oscillations governed by maximal mixing, $\theta_{23} \sim \pi/4$:
the relationship between the mass eigenstate $\nu_2$ and $\nu_3$
and the flavor eigenstates $\nu_\mu$ and $\nu_\tau$ involves a
rotation of 45 degrees. The magnitude of the mass difference
$\delta m_{23} = m_3^2-m_2^2$ between the two equal components
making up the flavor eigenstates is $\sim 2 \times 10^{-3}$
eV$^2$.  When I was written, the favored solution to the solar
neutrino problem was also neutrino mixing, but was described by a
much smaller mixing angle $\theta_{12}$ specifying the
relationship between mass eigenstates $\nu_1$ and $\nu_2$ and the
flavor eigenstates $\nu_e$ and $\nu_\mu$. The effects of this
small mixing are magnified by matter effects in the sun. This is
the so-called MSW effect and is described in I. But the results of
the Sudbury Neutrino Observatory provided a bit of a surprise.

\begin{figure*}
\begin{center}
\includegraphics[width=4.0in]{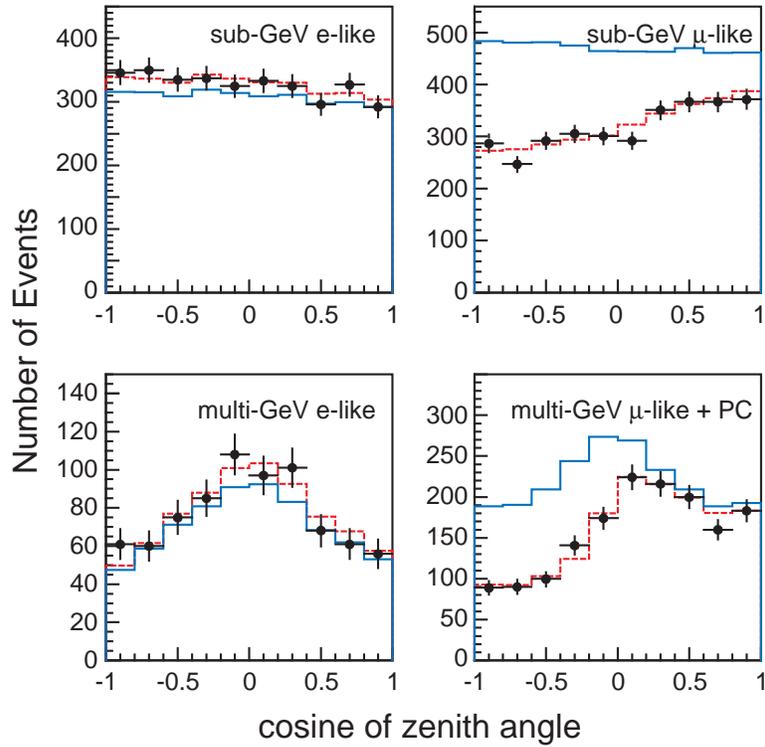}
\caption{\label{fig:atmos} The SuperKamiokande atmospheric
neutrino results showing excellent agreement between the predicted
(blue lines) and observed electron-like events, but a sharp
depletion in the muon-like events for neutrinos coming from below,
through the earth.  The results are fit very well by the
assumption of $\nu_\mu \rightarrow \nu_\tau$ oscillations with
maximal mixing (red lines).}
\end{center}
\end{figure*}

\section{The SNO Experiment}

Because both charged and neutral currents contribute to the
reaction important to SuperKamiokande,
\begin{equation}
\nu_x+e^-\rightarrow \nu_x+e^-,
\end{equation}
the experimentalists cannot easily distinguish $\nu_e$s from the
$\nu_\mu$s and $\nu_\tau$s: the detector records both fluxes,
though with a reduced sensitivity (0.15) for the heavy-flavor
types.  The reaction produces energetic recoil electrons which
generate Cerenkov radiation that is recorded in an array of
phototubes surrounding the detector.  As the cross section is
sharply forward peaked, the correlation with the position of the
sun can be used to ``cut'' background contributions associated
with cosmic rays and radioactivity in the rock walls surrounding
the detector. Because the threshold for electron detection is
$\sim 6$ MeV, only the high energy portion of the ${}^8$B solar
neutrino flux is measured. These are the same neutrinos that
dominate the radiochemical measurements of Ray Davis:
Superkamiokande confirmed that this flux was substantially below
that predicted by the standard solar model (SSM)\cite{bah}
\begin{equation}
\phi_{SSM}(\nu_x)=5.44\times 10^6\,{\rm cm}^{-2}{\rm
sec}^{-1}\label{eq:ssm}
\end{equation}
When the Davis and SuperKamiokande results were combined with
those from the gallium experiments SAGE and GALLEX, the resulting
three constraints on the three principal solar neutrino sources
($^8$B, $^7$Be, and the low-energy pp fluxes) produced a
surprising result. No combination of these fluxes could reproduce
the combined data well.  Though circumstantial, this evidence
indicated that the solution to the solar neutrino problem would
not be found in the SSM, but instead would require new particle
physics.

The favored explanation became neutrino oscillations which, as we
have just summarized, occur for massive neutrinos when the weak
and mass eigenstates do not coincide. The SuperKamiokande
discrepancy, a solar neutrino rate less than half that expected
from the SSM, would require that approximately two-thirds of the
high energy electron neutrinos generated in the solar interior
oscillate into other species -- $\nu_\mu,\nu_\tau$ -- before
reaching earth.  Low-energy $\nu_\mu$s and $\nu_\tau$s are
invisible to the Davis and SAGE/GALLEX detectors and scatter
electrons in the SuperKamiokande detector with a reduced cross
section, as we noted previously.

The key idea behind the Sudbury Neutrino Observatory (SNO) was
construction of a detector that would have multiple detection
channels, recording the $\nu_e$s by one reaction and the total
flux of all neutrinos ($\nu_e$s + $\nu_\mu$s + $\nu_\tau$s) by
another.  This was accomplished by replacing the ordinary water in
a water Cerenkov detector with heavy water -- D$_2$O instead of
H$_2$O. The charged-current (CC) channel that records the $\nu_e$s
is analogous to the reaction used in the Davis detector
\begin{equation}
\nu_e+d\rightarrow p+p+e^-.
\end{equation}
As the electron produced in this reaction carries off most of the
neutrino energy, its detection in the SNO detector (by the
Cerenkov light it generates) allows the
experimentalists to determine the spectrum of solar $\nu_e$s, not just
the flux.  A second reaction, the neutral current (NC) breakup of deuterium,
gives the total flux, independent of flavor (the $\nu_e$, $\nu_\mu$,
and $\nu_\tau$ cross sections are identical),
\begin{equation}
\nu_x+d\rightarrow n+p+\nu_x.
\end{equation}
The only signal for this reaction in a water Cerenkov detector is
the neutron, which can be observed as it captures via the
$(n,\gamma)$ reaction.  SNO is currently operating with salt added
to the water, as Cl in the salt is an excellent $(n,\gamma)$
target, producing about 8 MeV in $\gamma$s.

While the strategy may sound straightforward, producing such a
detector was an enormous undertaking.  The needed heavy water --
worth about \$300M -- was available through the Canadian
government because of its CANDU reactor program.  The
single-neutron detection required for the neutral current reaction
is possible only if backgrounds are extremely low.  For this
reason the detector had to be placed very deep underground,
beneath approximately two kilometers of rock, so that cosmic-ray
muon backgrounds would be reduced to less than 1\% of that found
in the SuperKamiokande detector.  The experimentalists found the
needed site in an active nickel mine, the Sudbury mine in Ontario,
Canada, where they worked with the miners to carve out a
10-story-high cavity on the mine's 6800-ft level.  Trace
quantities of radioactivity were another background concern: if a
thimblefull of dust were introduced into the massive cavity during
construction, the resulting neutrons from U and Th could cause the
experiment to fail.  Thus, despite the mining activities that
continued around them, the experimentalists constructed their
detector to the strictest cleanroom standards.  The detector also
provided a third detection channel, neutrino elastic scattering
(ES) off electrons (Eq. (1)), which we have noted is sensitive to
$\nu_e$s and, with reduced sensitivity, $\nu_\mu$s and
$\nu_\tau$s.

The ES reaction, of course, provides SNO a direct
cross check against SuperKamiokande.  SNO's threshold for measuring
these electrons is about 5 MeV. Assuming no oscillations, SNO's
detection rate is equivalent to a $\nu_e$ flux of
\begin{equation}
\phi_{SNO}^{ES}= 2.39\pm 0.34({\rm stat})\pm 0.15({\rm syst})
\times 10^{6}\,{\rm cm}^{-2}{\rm sec}^{-1},
\end{equation}
a result in excellent accord with that from SuperKamiokande,
\begin{equation}
\phi_{SK}^{ES}=2.32\pm 0.03({\rm stat})\pm 0.06({\rm syst})
\times 10^{6}\,{\rm cm}^{-2}{\rm sec}^{-1}.
\end{equation}
The greater accuracy of the SuperKamiokande result reflects the
larger mass (50 kilotons) and longer running time of the Japanese
experiment.  (SNO contains, in addition to the one kiloton of
heavy water in its central acrylic vessel, an additional seven
kilotons of ordinary water that surrounds the central vessel,
helping to shield it.)

The crucial new information provided by SNO comes from the two
reactions on deuterium.  The CC current channel is only sensitive
to $\nu_e$s.  Under the assumption of an undistorted $^8$B
neutrino flux, SNO experimentalists deduced
\begin{equation}
\phi_{SNO}^{CC}(\nu_e)=1.75\pm 0.07({\rm stat})\pm 0.12({\rm sys})\pm 0.05({\rm
theory})
\times 10^{6}\,{\rm cm}^{-2}{\rm sec}^{-1}.
\end{equation}
The CC flux is less than that deduced from the ES rate, indicating
that $\nu_\mu$s and $\nu_\tau$s must be contributing to the
later. From the difference between the SuperKamiokande ES and the
SNO CC results
\begin{equation}
\delta\phi=0.57\pm 0.17\times 10^{6}\,{\rm cm}^{-2}{\rm sec}^{-1}
\end{equation}
and recalling that the $\nu_\mu/\nu_\tau$ ES cross
section is only 0.15 that for the $\nu_e$, one deduces
the heavy-flavor contribution to the solar neutrino flux
\begin{equation}
\phi(\nu_\mu/\nu_\tau)=3.69\pm 1.13\times 10^{6}\,{\rm cm}^{-2}{\rm sec}^{-1}.
\end{equation}
That is, approximately two-thirds of the solar neutrino flux is
in these flavors.

While the first SNO analysis was done in the manner described
above, a second publication gave the long awaited NC results. This
allowed a direct and very accurate determination of the flavor
content of solar neutrinos, without the need for combining results
from two experiments. The published NC results were obtained
without the addition of salt to the detector: the neutron was
identified by the 6.25 MeV $\gamma$ ray it produces by capturing
on deuterium.  The resulting total flux, independent of flavor, is
\begin{equation}
\phi_{SNO}^{NC}(\nu_x)=5.09\pm 0.44({\rm stat})\pm 0.45({\rm syst})
\times 10^{6}\,{\rm cm}^{-2}{\rm sec}^{-1}.
\end{equation}
Combining with the CC signal yields
\begin{eqnarray}
\phi_{SNO}(\nu_e)=1.76\pm 0.05({\rm stat})\pm 0.09({\rm syst})
\times 10^{6}\,{\rm cm}^{-2}{\rm sec}^{-1} \nonumber \\
\phi_{SNO}(\nu_\mu/\nu_\tau)=3.41\pm 0.45({\rm stat})\pm 0.46({\rm syst})
\times 10^{6}\,{\rm cm}^{-2}{\rm sec}^{-1}
\end{eqnarray}
The presence of heavy-flavor solar neutrinos and thus neutrino
oscillations is confirmed at the 5.3$\sigma$ level!  Furthermore
the total flux is in excellent agreement with the predictions of
the SSM---Eq. \ref{eq:ssm}, an important vindication of stellar
evolution theory.

The SNO analysis is summarized in Figure 2, which shows the three
bands corresponding to the CC, NC, and ES measurements coinciding
in a single region.  These results can now be combined with other
solar neutrino measurements to determine the parameters -- the
mixing angle and mass-squared difference -- governing the
oscillations.  At the time I was written, there were several
contending solutions, though the data favored one characterized by
a small mixing angle (thus called the SMA solution).  Figure 3
shows that the SNO result has determined an oscillation solution
that, at 99\% confidence level, is unique -- and as in the
atmospheric neutrino case, it has a large mixing angle,
$\theta_{12} \sim 30$ degrees. This LMA oscillation is clearly
distinct from that seen with atmospheric neutrinos, with $\delta
m_{12}^2 = m_2^2 - m_1^2$ centered on a region $\sim 8 \times
10^{-5}$ eV$^2$.

The discovery that the atmospheric and solar neutrino problems are
both due to neutrino oscillations has provided the first evidence
for physics beyond the standard model.  That neutrinos provided
this evidence is perhaps not unexpected: if the standard model is
viewed as an effective theory, one largely valid in our low-energy
world but missing physics relevant to very high energies, beyond
the reach of current accelerators, then a neutrino mass term is
the lowest-order correction that can be added to that theory.  But
a surprise is the large mixing angles characterizing the neutrino
oscillations -- which contradicts the simple prejudice that
neutrino mixing angles might be similar to the small angles
familiar from quark mixing.  Perhaps this simply reinforces
something that should have been apparent at the outset: with their
small masses and distinctive mixings, neutrinos likely have an
underlying mechanism for mass generation that differs from that of
the other standard model fermions.

\begin{figure}
\begin{center}
\includegraphics[width=4.9in]{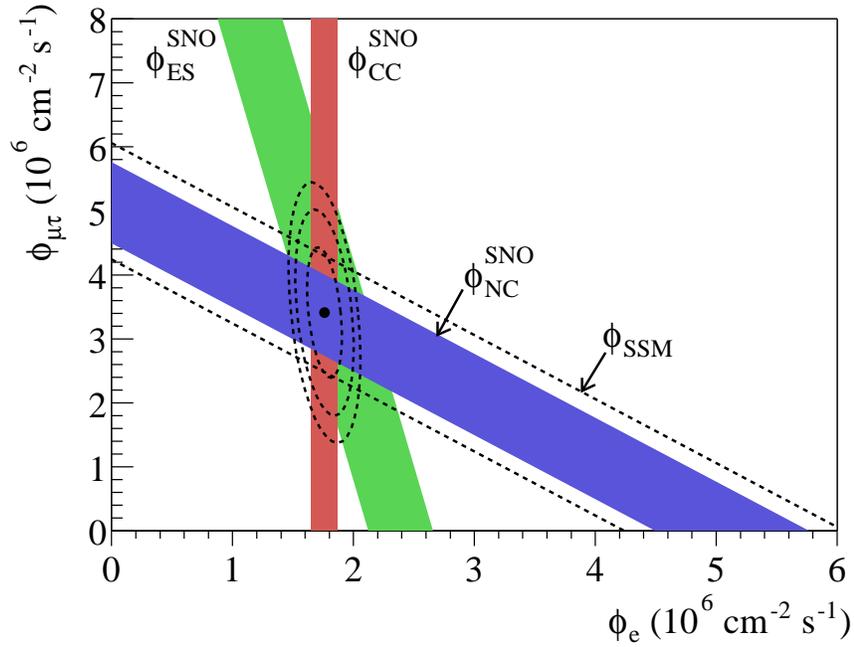}
\caption{\label{hime_plot}Flux of ${}^{8}$B solar neutrinos is
divided into $\nu_\mu/\nu_\tau$ and $\nu_e$ flavors by the SNO analysis.
The diagonal bands show the total ${}^{8}$B flux as predicted by the
SSM (dashed lines) and that
 measured with the NC reaction in SNO (solid band).  The widths of
these bands
 represent the $\pm 1\sigma$ errors.  The bands intersect in a single region
 for $\phi(\nu_e)$ and $\phi(\nu_\mu/\nu_\tau)$, indicating that the combined
flux results are consistent
 with neutrino flavor transformation assuming no distortion in the
${}^{8}$B neutrino energy spectrum.}
\end{center}
\end{figure}

\begin{figure}
\begin{center}
\includegraphics[width=3.9in]{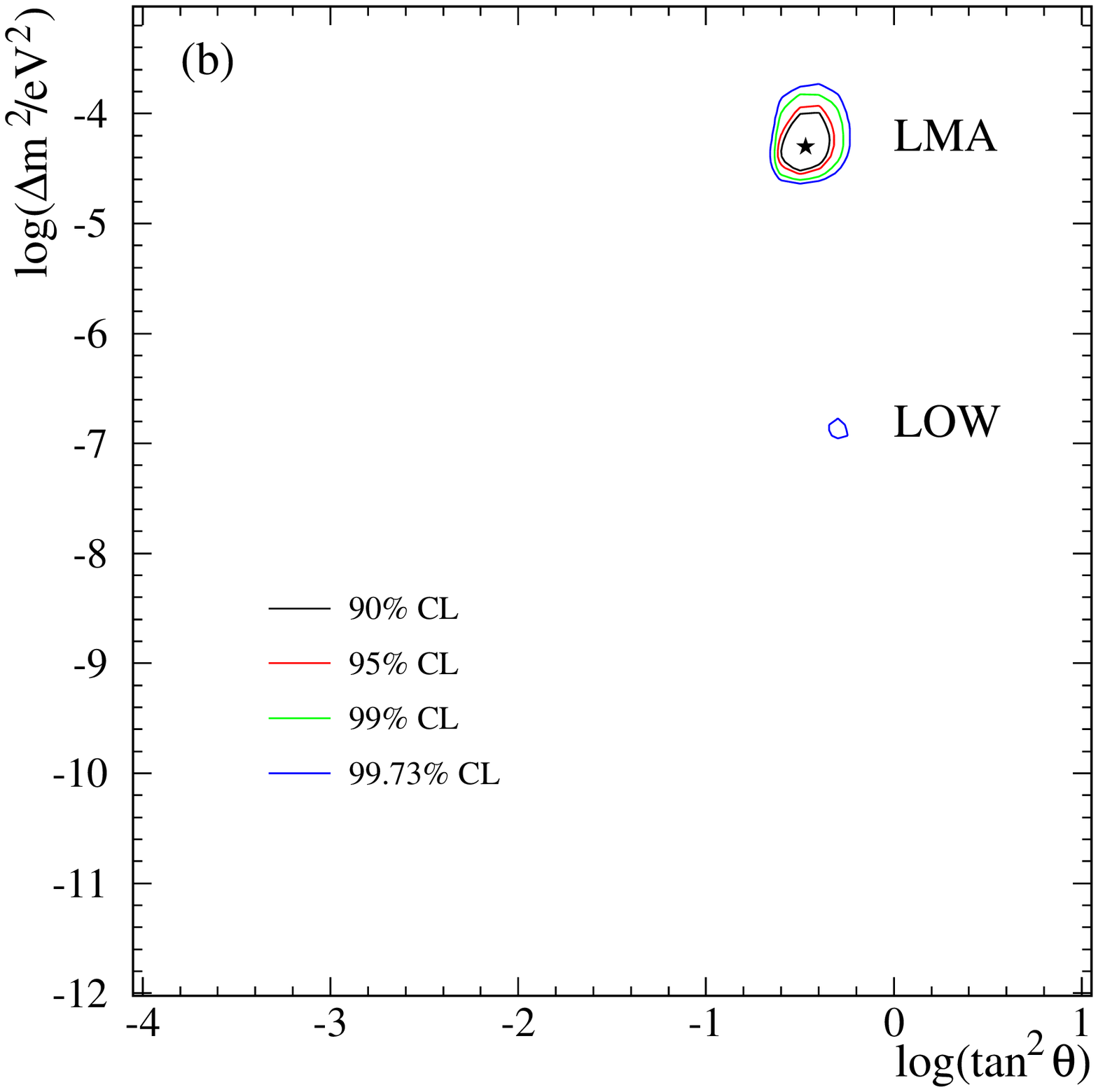}
\caption{\label{fig:hime_plot} For two-flavor mixing, the values
of the mass difference and mixing angle consistent with the
world's data on solar neutrinos, post SNO.  At 99\%
confidence level the addition of SNO data isolates a unique, large-mixing angle
solution.}
\end{center}
\end{figure}

\section{The KamLAND Experiment}

One remarkable aspect of the solar and atmospheric neutrino
discoveries is that the derived oscillation parameters are
within the reach of terrestrial experiments.  This did not
have to be the case -- solar neutrinos are sensitive to
neutrino mass differences as small as 10$^{-12}$ eV$^2$,
for which terrestrial experiments would be unthinkable.

The first terrestrial experiment to probe solar neutrino
oscillation parameters, KamLAND, very recently reported first
results.  The acronym KamLAND stands for Kamioka Liquid
scintillator Anti-Neutrino Detector.  The inner detector consists
of one kiloton of liquid scintillator contained in a spherical
balloon, 13m in diameter.  The balloon is suspended in the old
Kamioka cavity (where SuperKamiokande's predecessor was housed) by
Kevlar ropes, with the region between the balloon and an
18m-diameter stainless steel containment vessel filled with
additional scintillator (serving to shield the target from
external radiation).  Several Japanese power reactors are about
180 km from the Kamioka site, and the electron antineutrinos
emitted by nuclear reactions in the cores of these reactors can be
detected in KamLAND via the inverse beta decay reaction
\begin{equation}
\bar{\nu}_e+p\rightarrow e^++n,
\end{equation}
where the $e^+$ is seen in coincidence with the delayed
2.2 MeV $\gamma$ ray produced by the capture of the accompanying
neutron on a proton.  This
coincidence allows the experimentalists to
distinguish $\bar{\nu}_e$ reactions from background.

>From the reactor operations records, which
the power companies have made available, KamLAND experimentalists
can calculate the resulting flux at Kamiokande to a precision of $\sim 2$\%,
in the absence of oscillations.  Thus, if a significant fraction of the
reactor $\bar{\nu}_e$s oscillate into $\bar{\nu}_\mu$s or
$\bar{\nu}_\tau$s before reaching the detector, a low rate of
$e^+$/capture-$\gamma$-ray coincidences will be evident: this
is an example of the ``disappearance'' oscillation technique we
described in I. For the 162 ton/yr exposure so far
reported by the KamLAND collaboration, the
number of events expected in the absence of oscillations is $86.8\pm 5.6$.
But the number measured is 54 -- just 61\% of the no-oscillation
expectation.  From the two-neutrino-flavors oscillation
survival probability
\begin{equation}
P(\bar{\nu}_e\rightarrow\bar{\nu}_e)\simeq
1-\sin^22\theta_{12}\sin^2{\delta m_{12}^2L\over 4E_\nu}
\end{equation}
one obtains the oscillation parameters of Figure 4.  KamLAND
confirms the LMA solution and significantly narrows SNO's allowed
region (the red area in Figure 4).  KamLAND has excellent sensitivity
to $\delta m_{12}^2$ but less sensitivity to $\sin^2 2\theta_{12}$ (due
to uncertainties in the shape of the reactor $\bar{\nu}_e$ spectrum).
The result is the separation of the SNO LMA allowed region into
two parts, with the best-fit $\delta m_{12}^2 \sim 7 \times 10^{-5}$ eV$^2$,
but with a larger mass difference $\sim 1.5 \times 10^{-4}$ eV$^2$ also
fitting well.  KamLAND is an excellent example of complementary
terrestrial and astrophysical measurements: solar neutrino experiments
provide our best constraints on $\theta_{12}$, but KamLAND places the
tightest bounds on $\delta m_{12}^2$.

\begin{figure}[!ht]
\centerline{\psfig{figure=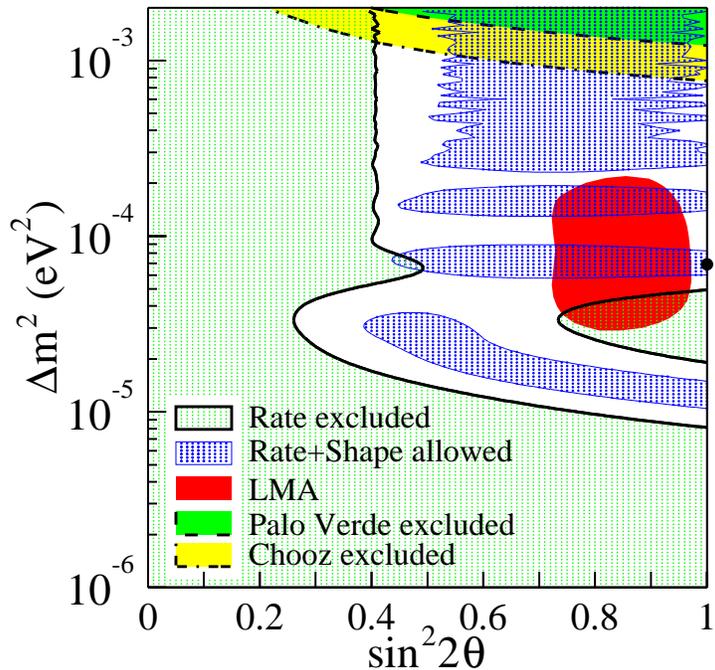,width=3.8in}}
\caption{The 95\% c.l. LMA allowed region of SNO and other solar
neutrino experiments is shown in red.  The regions marked
``Rate and Shape allowed'' show the 95\% c.l. KamLAND allowed solutions.
The thick dot indicates the best fit to the KamLAND data,
corresponding to $\sin^2 2\theta_{12} \sim 1.0$ and
$\delta m_{12}^2 \sim 6.9 \times 10^{-5}$ eV$^2$.}
\end{figure}

\section{WMAP, Double Beta Decay, and Neutrino Mass}

Despite the wonderful recent discoveries in neutrino physics,
there remain quite a number of open questions.  Several have to do
with the pattern of neutrino masses. All of the experiments
described above probe mass {\it differences}, not absolute
neutrino masses. Furthermore the atmospheric neutrino experiments
only constrain the magnitude of $\delta m_{23}^2$, and not its
sign. As a result, there exist several mass patterns fully
consistent with all known data.  One choice would assign $m_3$ to
be the heaviest neutrino, split by the atmospheric mass difference
$\delta m_{23}^2 \sim 2 \times 10^{-3}$ eV$^2$ from a lighter,
nearly degenerate pair of neutrinos responsible for solar neutrino
oscillations.  (This pair is split by solar neutrino mass
difference $\delta m_{12}^2 \sim 10^{-5}$ eV$^2$.)  However, as
the sign of $\delta m_{23}^2$ is not known, it is also possible
that $m_3$ is the lightest neutrino, with the nearly degenerate
$m_1$ and $m_2$ heavier. Finally, the best direct laboratory
constraint on absolute neutrino masses comes from studies of
tritium beta decay, as described in I.  Studies of the tritium
spectrum near its endpoint energy places a bound of 2.2 eV
\cite{2pt2} on the $\bar{\nu}_e$ mass (or more properly, on the
principal mass eigenstate contributing to the $\bar{\nu}_e$).
Consequently, one can add an overall scale of up to 2.2 eV to the
mass splittings described above.  That is, no terrestrial
measurement rules out three nearly degenerate neutrinos, each with
a mass $\sim$ 2.2 eV, but split by requisite $\delta m_{atmos}^2$
and $\delta m_{solar}^2$.

As discussed in I, the absolute neutrino mass is crucial in
cosmology, as a sea of neutrinos produced in the Big Bang pervades
all of space. If these neutrinos carry a significant mass, they
would constitute an important component of particle dark matter,
invisibly affecting the structure and expansion of our universe.
Light neutrinos, such as those we have been discussing, decouple
from the rest of the matter as relativistic particles, about one
second after the Big Bang. Their number density and temperature
can be calculated at the time of decoupling and today.  Their
contribution to the universe's mass-energy budget is now dominated
by their masses,
\begin{equation}
\rho_\nu = 0.022 \rho_{crit} \sum_i {m_\nu(i) \over {\rm eV}},
\end{equation}
where $\rho_{crit}$ is the critical density that will just close
the universe.  A variety of cosmological probes suggest that our
universe is very close to the critical density---$\rho/\rho_{crit}
= 1.0 \pm 0.04$.

>From the discussion above, we know that at least one neutrino must have
a mass of at least $\sqrt{\delta m^2_{atmos}}$.  Similarly, the tritium
beta decay limit places an upper bound on the sum of the masses of 6.6 eV
(corresponding to three nearly generate neutrinos of mass 2.2 eV).
It follows that the neutrino contribution to dark matter is bounded above
and below
\begin{equation}
 0.0012 \lsim \rho_\nu/\rho_{crit} \lsim 0.15.
\end{equation}
This broad range implies that the amount of mass in neutrinos
could easily exceed all the familiar baryon matter -- stars, dust,
gas clouds, and us -- visible or invisible: big-bang
nucleosynthesis and precision measurements of the cosmic microwave
background (CMB) both indicate that $\rho_{baryons}/\rho_{crit} \sim
0.042$.

However, in the past few years a series of extraordinarily precise measurements
have been made in cosmology.  One of these is the recent WMAP full-sky map of the
CMB and its subtle (few millionths of a degree)
temperature anisotropies.  This is the oldest
light in the universe, the photons that decoupled from matter at the time atoms
formed,
about 380,000 years after the Big Bang.  The CMB temperature anisotropies tell us
about the structure of the universe -- its clumpiness -- at this very early
epoch.  Measurements of distant SNIa supernovae -- a sort of ``standard candle''
by which astronomers can measure cosmological distances -- have constrained the
expansion rate and mass/energy budget of the universe.  Large-scale surveys, such
as the 2dF Galaxy Redshift Survey, have mapped the distribution of visible
matter in the universe today and in recent times.  The result from combining these
and other cosmological probes is a rather sharp constraint on the amount of
hot dark matter -- in particular, the mass density in neutrinos -- that can be
allowed, given that our universe has evolved to its present state.  One finds
\begin{equation}
 \rho_\nu/\rho_{crit} \lsim 0.022,
\end{equation}
that is, an upper bound on the sum of neutrino masses of about 1.0 eV.  (Some
analyses claim even tighter upper bounds.)  This bound is significantly tighter than
that of Eq. (14), derived from laboratory data only.

Cosmology now demands that neutrino dark matter can make
up no more than 2-3\% of the universe's mass-energy budget and, in particular, is
less important than other forms of matter we know about (e.g., nucleons).
(We will not go into the disconcerting fact that at least 93\% of the universe's
mass-energy
budget appears to be dark energy and cold dark matter that we have not yet
adequately
characterized!)  It also tells us that neutrino mass at the $\sim$ 1 eV level now
effects cosmological analyses: such analyses would constrain other cosmological
parameters more tightly if neutrino masses were
measured, rather than being free parameters that one must dial in cosmological
models until unacceptable deviations are found.  (The one eV bound was derived in
this way.)  This underscores how important it is to significantly improve
laboratory mass limits.

One possibility is a new-generation tritium experiment: there is a
serious effort underway to improve the current bound on the
$\bar{\nu}_e$ mass to about 0.3 eV, which would then place an
upper bound on the sum of the masses of about 0.9 eV \cite{2pt2}.
Another possibility -- less definitive, perhaps, but with even
greater reach -- is offered by new-generation neutrinoless double
beta decay experiments.

The phenomenon of neutrinoless double beta decay, described in I,
tests not only mass, but also whether a standard model symmetry
called lepton number conservation is violated.  In neutrinoless
double beta decay a nucleus spontaneously decays by changing its
charge by two units while emitting two electrons,
\begin{equation}
(A,Z) \rightarrow (A,Z+1)+e^-+\bar{\nu}_e \rightarrow (A,Z+2)+e^-+e^-,
\end{equation}
where the intermediate nuclear state $(A,Z+1)$ is virtual. The
emitted electrons carry off the entire nuclear energy release,
allowing this process to be distinguished from the
standard-model-allowed process of two-neutrino double beta decay
(where the energy is shared between two electrons and two
$\bar{\nu}_e$s in the final state). The neutrinoless process
clearly violates lepton number, as two leptons (the electrons)
are spontaneously produced.  (By contrast, as the $e^-$ carries $l
= +1$ and the $\bar{\nu}_e$ $l=-1$, two-neutrino double beta decay
conserves lepton number.)

What conditions will lead to neutrinoless double beta decay?  The
necessary lepton number violation is present if the neutrino is a
Majorana particle---{\it i.e.}, is identical to its antiparticle.
Most theoretical models include Majorana neutrinos: this is part
of the mechanism that allows us to understand why neutrinos have
masses much smaller than those of the other standard-model
fermions, such as electrons and quarks.  But the existence of a
Majorana neutrino alone is not sufficient because of the exact
handedness of neutrinos, which we discussed in I.  In the
neutrinoless double beta decay reaction written above, the
$\bar{\nu}_e$ appearing in the intermediate nuclear state was
produced in the nucleus by the neutron $\beta$ decay reaction $n
\rightarrow p + e^- + \bar{\nu}_e$. To complete the decay, the
antineutrino must be reabsorbed on a second neutron, $\nu_e + n
\rightarrow p + e^-$. At first glance, if $\bar{\nu}_e = \nu_e$,
{\it i.e.}, if the neutrino is its own antiparticle, this
reabsorption looks possible. However, this conclusion overlooks
the neutrino handedness. In the first step, the $\bar{\nu}_e$
produced is righthanded, while the second reaction only proceeds
if the $\nu_e$ is lefthanded.  Thus it would appear that the
neutrinoless process is forbidden, even if $\bar{\nu}_e = \nu_e$.

This argument, however, overlooks the effects of neutrino mass: a small neutrino
mass breaks the exact neutrino handedness, allowing neutrinoless $\beta \beta$ decay
to proceed, though the amplitude is suppressed by the factor $m_\nu/E_\nu$, where $E_\nu \sim 30$ MeV
is the typical energy of the exchanged neutrino.  It follows that neutrinoless
double beta decay measures the neutrino mass -- at least the Majorana portion
of that mass.  (Making this statement more precise, unfortunately, takes us beyond
the limits of this paper.)

In the simplest case -- a single Majorana mass eigenstate
dominating the $\beta\beta$ decay -- the neutrinoless amplitude is
proportional to $U_{ei}^2 m_i$, where $U_{ei}^2$ is the mixing
probability of the $i$th mass eigenstate in the $\nu_e$ and $m_i$
is the mass.  Currently the best neutrinoless $\beta\beta$ decay
limits are those obtained by the Heidelberg-Moscow and IGEX
enriched (86\%) $^{76}$Ge experiments, which are both probing
lifetimes beyond $10^{25}$ years -- corresponding to roughly one
decay per kg-year!  These experiments employ Ge crystals -- the Ge
is both source and detector -- containing about 10 kg of active
material.  Next-generation experiments, using a variety of double
beta decay sources ($^{76}$Ge, $^{136}$Xe, $^{100}$Mo), have been
proposed at the one ton scale.  These have as their goals
sensitivities to neutrino masses of 10-50 milli-eV, corresponding
to lifetimes well in excess of $10^{27}$ y.  One important
motivation for these heroic proposals is the number $\sqrt{\delta
m_{atmos}^2} \sim$ 55 milli-eV: in several scenarios accommodating
the solar and atmospheric oscillation results, this scale plays a
role in determining the level at which neutrinoless $\beta\beta$
decay might be observed.

A few members of the Heidelberg-Moscow collaboration have claimed
that their present results are not a limit, but rather a detection
of neutrinoless $\beta\beta$ decay, with a best value for the
Majorana neutrino mass of $\sim$ 0.4 eV \cite{dbd}.  This claim
has also been strongly criticized by a group that argues that the
claimed peak, shown in Figure 5, is not statistically significant.
Regardless, this claim will clearly be tested soon, in other
$\beta\beta$ decay experiments and in future cosmological tests,
which promise to soon be probing masses $\sim$ 0.3 eV.

\begin{figure}
\begin{center}
\includegraphics[scale=0.40]
{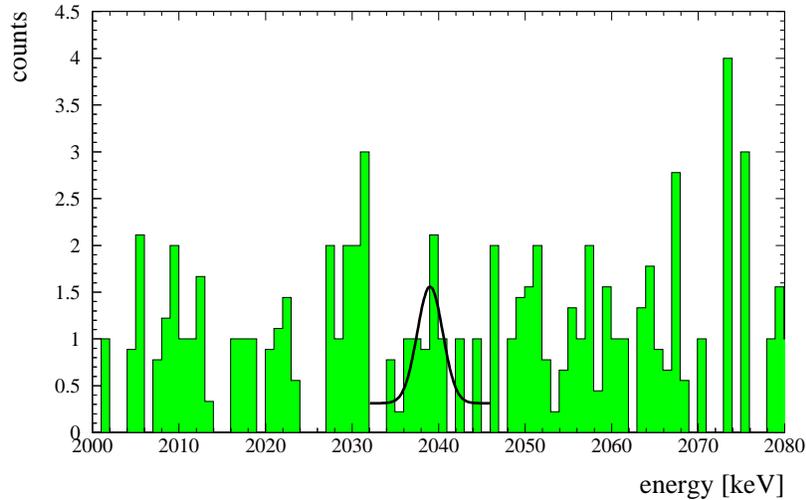} \caption{Spectrum found in ref. \cite{dbd}. The
claimed signal is as shown.}
\end{center}
\end{figure}

\section{Conclusion}

In the three years since the publication of I, several very significant
neutrino discoveries have been made:

\begin{itemize}
\item [i)]  SNO has shown that approximately two-thirds of the $^8$B
neutrinos that arrive on earth have oscillated into $\nu_\mu$s or
$\nu_\tau$s, thus demonstrating that new neutrino physics is
responsible for the solar neutrino puzzle first uncovered by Ray
Davis, Jr.  Together with the atmospheric neutrino discoveries of
SuperKamiokande, this discovery of an effect requiring massive
neutrinos and neutrino mixing is the first evidence for physics
beyond the standard model.  The SNO results for the total solar
neutrino flux, independent of flavor, are in excellent agreement
with the predictions of the standard model -- despite the
challenge of calculating a flux that varies as $T_c^{22}$(!),
where $T_c$ is the solar core temperature.  The SNO results, when
added to other solar neutrino data, isolate a single oscillation
scenario, the LMA solution.

\item [ii)]  The first terrestrial experiment to probe solar
neutrino oscillation parameters, KamLAND, has confirmed the
SNO results and further narrowed the LMA range of allowed $\delta m_{12}^2$.

\item [iii)]  Both the absolute scale of neutrino masses and the
detailed pattern of the masses remain unknown, as the results in
hand measure only mass differences (and leave the sign of
$\delta m_{23}^2$ undetermined).  The most stringent current bound
on the absolute scale of neutrino mass comes from recent precision
tests of cosmology (notably WMAP and the 2dF Galaxy Redshift Survey).
This limit, a bound of about 1 eV for the sum of neutrino masses,
is likely to improve as new surveys are done.  In addition, much
improved tritium $\beta$ decay and neutrinoless $\beta\beta$ decay
experiments are being planned.  There is one controversial claim
of an observation of neutrinoless $\beta\beta$ decay that must be
checked soon.
\end{itemize}

We stress, as we did in I, that this field is producing many new
results that promise to impact physics broadly.  The most common
mechanisms for explaining neutrino mass suggest that current
experiments are connected with phenomena far outside the standard
model, residing near the Grand Unified energy scale of $10^{16}$
GeV. Thus there is hope that, by fully determining the properties
of neutrinos -- a few of the unresolved problems have been
mentioned here -- we may equip theorists to begin constructing the
next standard model.  Neutrino physics is also crucial to
astrophysics -- not just the standard solar model, but also in
supernovae and in high-energy astrophysical environments -- and to
cosmology.  We now have identified the first component of particle
dark matter, though the significance of neutrino mass is still
unclear due to our ignorance of the overall scale.  Neutrinos
could prove central to one of cosmology's deepest questions, why
our universe is matter dominated, rather than matter-antimatter
symmetric.  But this is a story for another paper and another
time.

\begin{center}
{\bf Acknowledgement}
\end{center}
The work was supported in part by the U.S. Department of Energy
and by the National Science Foundation under award PHY-98-01875.


\begin{thebibliography}{99}
\bibitem{nui} W.C. Haxton and B.R. Holstein,``Neutrino physics,''
Am. J. Phys. {\bf 68}, 15-32 (2000).
\bibitem{dav} See information, including the Nobel lectures,
at the website of the Nobel Foundation---http://www.kva.se.
\bibitem{sno1} Q.R. Ahmad et al., ``Measurement of the rate of
$\nu_e + d \rightarrow p + p + e^-$ interactions produced by $^8$B
solar neutrinos at the Sudbury Neutrino Observatory,'' Phys. Rev.
Lett. {\bf 87}, 071301 (2001) pp 1-6.
\bibitem{sno2} Q.R. Ahmad et al., ``Direct evidence for neutrino
flavor transformation from neutral-current interactions in the
Sudbury Neutrino Observatory,''  Phys. Rev. Lett. {\bf 89}, 011301
(2002) pp 1-6.
\bibitem{kam} K. Eguchi et al., ``First results from KamLAND:
Evidence for reactor antineutrino disappearance,'' Phys. Rev.
Lett. {\bf 90}, 021802 (2003) pp 1-6.
\bibitem{wmap} D. N. Spergel et al., ``First year Wilkinson Microwave
Anisotropy Probe (WMAP) observations: Determination of
cosmological parameters,'' [arXiv: astro-ph/0302213] (submitted to
Ap. J.).
\bibitem{2d} J. Peacock et al., ``A measurement of the cosmological
mass density from clustering in the 2dF Galaxy Redshift Survey,''
Nature {\bf 410}, 169-73 (2001).
\bibitem{dbd} H.V. Klapdor-Kleingrothaus et al., ``Evidence for
neutrinoless double beta decay,'' Mod. Phys. Lett. A{\bf 16},
2409-20 (2001); C. E. Aaalseth et al., ``Comment on Evidence for
neutrinoless double beta decay,'' Mod. Phys. Lett. A{\bf 17},
1475-78 (2002).
\bibitem{dgh} See, {\it e.g.}, J.F. Donoghue, E. Golowich, and
B.R. Holstein {\bf Dynamics of the Standard Model}, Cambridge
University Press, New York (1992).
\bibitem{sol} R. Davis, Jr., D.S. Harmer, and K.C. Hoffman,
``Search for neutrinos from the sun,'' Phys. Rev. Lett. {\bf 20},
1205-12 (1968); J. N. Bahcall and R. Davis, Jr., ``Solar
neutrinos: a scientific puzzle,'' Science {\bf 191}, 264-67
(1976).
\bibitem{atm}Y. Fukuda et al., ``Evidence for the oscillation
of atmospheric neutrinos,'' Phys. Rev. Lett. {\bf 81}, 1562-65 (1998).
\bibitem{bah} J.N. Bahcall, M.H. Pinsonneault, and S. Basu, Ap. J.
{\bf 555}, 990-1012 (2001).
\bibitem{2pt2} J. F. Wilkerson and R. G. H. Robertson, ``Direct
measurements of neutrino mass,'' in {\bf Current Aspects of
Neutrino Physics} (Springer-Verlag, Berlin, 2001), ed. D. O.
Caldwell, p. 39-64.
\end{thebibliography}
\end{document}